\input harvmac\skip0=\baselineskip

\def\e{\epsilon}
\def\al{\alpha'}

\def\a{{\alpha}}
\def\b{\beta}
\def\G{\Gamma}
 
\def\g{\gamma}
\def\s{\sigma}
\def\th{\theta}
\def\k{\kappa}
\def\slf{F \llap{/}}
\def\Th{\Theta}
\def\d{\delta}

\lref\spin{T. Kugo, P. Townsend, ``Supersymmetry and the division
algebras'', Nucl. Phys B221 : 357 (1983), and Polchinski vol II,
appendix.} \lref\gsy{Davide Gaiotto, Andrew Strominger, Xi Yin,
``From AdS(3)/CFT(2) to black holes/topological
strings''[hep-th/0602046]} \lref\bw{Chris Beasley, Edward Witten,
``New instanton effects in string theory'', JHEP 0602:060 (2006)
[hep-th/0512039]} \lref\colemanlee{S. Coleman, K. Lee, ``
Escape from the menace of the giant wormholes'', Phys. Lett. B221:
242 (1989)} \lref\entropy{L. Susskind, J. Uglum, ``Blackhole Entropy
in canonical quantum gravity and superstring theory'', Phys. Rev. D50:
2700-2711 (1994) [hep-th/9401070]; M. Banados,
C. Teitelboim, J. Zanelli, ``Black hole entropy and
the dimensional continuation of the Gauss-Bonnet theorem'',
Phys. Rev. Lett. 72: 957-960 (1994) [gr-qc/9309026];
S. Carlip, C. Teitelboim, ``The offshell black hole'',
Class. Quant. Grav. 12: 1699-1704 (1995) [gr-qc/9312002]  } \lref\berk{N.
Berkovits, M. Bershadsky, T. Hauer, S. Zhukov, B. Zwiebach,
``Superstring theory on AdS(2) x S**2 as a coset supermanifold'',
Nucl.Phys.B567:61-86 (2000) [hep-th/9907200]} \lref\msw{Juan M.
Maldacena, Andrew Strominger, Edward Witten, ``Black hole entropy in
M theory'', JHEP 9712:002 (1997) [hep-th/9711053]} \lref\osv{Hirosi
Ooguri, Andrew Strominger, Cumrun Vafa, ``Black hole attractors and
the topological string'', Phys.Rev.D70:106007 (2004)
[hep-th/0405146]} \lref\farey{Robbert Dijkgraaf, Juan M. Maldacena,
Gregory W. Moore, Erik P. Verlinde, ``A Black hole Farey tail''
[hep-th/0005003]} \lref\au{Cumrun Vafa, ``Two dimensional
Yang-Mills, black holes and topological strings'' [hep-th/0406058]}
\lref\ad{Atish Dabholkar, ``Exact counting of black hole
microstates'', Phys.Rev.Lett.94:241301 (2005) [hep-th/0409148]}
\lref\at{Mina Aganagic, Hirosi Ooguri, Natalia Saulina, Cumrun Vafa,
``Black holes, q-deformed 2d Yang-Mills, and non-perturbative
topological strings'', Nucl.Phys.B715:304-348 (2005)
[hep-th/0411280]} \lref\ap{Erik P. Verlinde, ``Attractors and the
holomorphic anomaly'' [hep-th/0412139]} \lref\ac{G. Lopes Cardoso,
B. de Wit, J. Kappeli, T. Mohaupt, ``Asymptotic degeneracy of dyonic
N = 4 string states and black hole entropy'', JHEP 0412:075 (2004)
[hep-th/0412287]} \lref\as{Atish Dabholkar, Frederik Denef, Gregory
W. Moore, Boris Pioline, ``Exact and asymptotic degeneracies of
small black holes'', JHEP 0508:021 (2005) [hep-th/0502157]}
\lref\ash{Hirosi Ooguri, Cumrun Vafa, Erik P. Verlinde,
``Hartle-Hawking wave-function for flux compactifications'',
Lett.Math.Phys.74:311-342 (2005) [hep-th/0502211]} \lref\ao{Mina
Aganagic, Andrew Neitzke, Cumrun Vafa, ``BPS microstates and the
open topological string wave function''[hep-th/0504054]}
\lref\an{Robbert Dijkgraaf, Rajesh Gopakumar, Hirosi Ooguri, Cumrun
Vafa, ``Baby universes in string theory'', Phys.Rev.D73:066002
(2006) [hep-th/0504221]} \lref\az{Ashoke Sen, ``Black hole entropy
function and the attractor mechanism in higher derivative gravity'',
JHEP 0509:038 (2005) [hep-th/0506177]} \lref\bu{Atish Dabholkar,
Frederik Denef, Gregory W. Moore, Boris Pioline, ``Precision
counting of small black holes'', JHEP 0510:096 (2005)
[hep-th/0507014]} \lref\bd{David Shih, Xi Yin, ``Exact black hole
degeneracies and the topological string'', JHEP 0604:034 (2006)
[hep-th/0508174]} \lref\bt{Mina Aganagic, Daniel Jafferis, Natalia
Saulina, ``Branes, black holes and topological strings on toric
Calabi-Yau manifolds'', [hep-th/0512245]} \lref\bp{Murat Gunaydin,
Andrew Neitzke, Boris Pioline, Andrew Waldron, ``BPS black holes,
quantum attractor flows and automorphic forms'', Phys.Rev.D73:084019
(2006) [hep-th/0512296]} \lref\bc{G. Lopes Cardoso, B. de Wit, J.
Kappeli, T. Mohaupt, ``Black hole partition functions and duality'',
JHEP 0603:074 (2006) [hep-th/0601108]} \lref\bs{Daniel Jafferis,
``Crystals and intersecting branes''
 [hep-th/0607032]}
\lref\bo{Davide Gaiotto, Andrew Strominger, Xi Yin, ``The M5-Brane Elliptic Genus: Modularity and BPS States''
[hep-th/0607010]}

\Title{\vbox{\baselineskip12pt }}{Why $Z_{\rm BH}=|Z_{\rm top}|^2$} \centerline{Chris
Beasley, Davide Gaiotto, Monica Guica, } \centerline{Lisa Huang,
Andrew Strominger and Xi Yin}
\smallskip
\centerline{Jefferson Physical Laboratory, Harvard University,
Cambridge, MA 02138} \vskip .6in \centerline{\bf Abstract} {It is
argued, using an M-theory lift, that the IIA partition function on
a euclidean $AdS_2\times S^2\times CY_3$ attractor geometry
computes the modified elliptic genus $Z_{BH}$ of the associated
black hole in a large charge expansion. The partition function is
then evaluated using the Green-Schwarz formalism. After localizing
the worldsheet path integral with the addition of an exact term,
contributions arise only from the center of $AdS_2$ and the north
and south poles of  $S^2$. These are the toplogical and
anti-topological string partition functions $Z_{top}$ and $\bar
Z_{top}$ respectively. We thereby directly reproduce the
perturbative relation $Z_{ BH}=|Z_{ top}|^2$.} \vskip .3in
\smallskip
\Date{August 2, 2006}

\listtoc \writetoc
\newsec{Introduction}
     Several years ago it was conjectured \osv\ that the modified elliptic genus $Z_{BH}$ of a IIA Calabi-Yau black hole and the topological string partition function $Z_{top}$ on the associated Calabi-Yau attractor enjoy a relationship of the form
     \eqn\dfto{Z_{BH}=|Z_{top}|^2}
  to all orders in  perturbation theory. Some evidence for and refinements of this conjecture have been presented in \refs{\au\ad\at\ap\ac\as\ash\ao\an\az\bu\bd\bt\bp\bc\bs-\bo}.

 The motivation given in \osv\ was simply a detailed comparison of the large-charge perturbation expansions of
 both sides of the relation \dfto.  In this comparison, deriving the exponent `2' appearing on the right hand side
 required detailed knowledge of the conventions involved. No insight was offered as to why the square should appear.
 Recently the origin of the square was explained \gsy\ by lifting to M-theory. At $low$ temperatures, the black hole
 partition function can be expressed as a dilute-gas sum over BPS wrapped membranes in $AdS_3\times S^2\times CY_3$.
 There are two types of such:  those wrapping holomorphic cycles and localized near the north pole, and those wrapping
 anti-holomorphic cycles and localized near the south pole.  These two types of wrapped branes were shown to give
 factors $Z_{top}$ and $\bar Z_{top}$ respectively.  This is not exactly the OSV relation \dfto\ which is a statement
 about $high$ temperatures, where the membranes are not dilute.  A modular transformation is then used to turn  this
 low-temperature  observation into the  high-temperature relation \dfto.

 It is of interest to find a direct derivation of \dfto\ which does not involve a modular transformation.
 This requires thinking about the M-theory partition function at high temperatures. At high temperatures  the  thermal circle is small, and M-theory reduces to IIA (because we want a $(-)^F$ insertion, the thermal circle has the correct periodic boundary conditions). This leads us back to the IIA theory on an $AdS_2\times S^2 \times CY_3$ attractor.  In this paper we will in fact argue, using worldsheet methods, that the
 perturbative IIA worldsheet partition function on an attractor gives precisely  $|Z_{top}|^2$, with one factor coming from worldsheet instantons at the north pole and the other from worldsheet anti-instantons at the south pole.

 This paper is organized as follows. In section 2 we consider the lift of the euclidean IIA attractor geometry to
M-theory.  The result is a quotient of $AdS_3\times S^2\times CY_3$
in which  the asymptotic $AdS_3$ boundary is a torus.  $AdS_3/CFT_2$
duality then implies
 (noting the fermion boundary conditions) that this path integral computes the (modified) elliptic genus
 for the black hole. Two notable features of this discussion is that it involves complex
 geometries and that, unlike in \farey, no ``Wilson line" insertions are required.  Having argued that the path
 integral on the IIA attractor computes the elliptic genus, we then proceed, in section 3, to a direct evaluation
 in string perturbation theory using the Green-Schwarz formalism.  An immediate difficulty is a factor
 of $\infty\times 0 $ coming from bosonic and fermionic worldsheet zero modes on $AdS_2\times S^2$.  To compute
 this factor we localize the path integral to the north and south poles of $S^2$ and the center of $AdS_2$ by adding an
 exact term to the action.  Relying heavily on the worldsheet computation of \bw\ for the internal $CY_3$ factor, we then
 find that the path integral gives precisely $|Z_{top}|^2$, as concluded in section 4. A number of technical points are detailed in the appendices.

\newsec{Euclidean IIA Attractors and the Elliptic Genus }
\subsec{The D4-D0 Attractor}

In this subsection we briefly describe the euclidean Calabi-Yau
attractor geometry for the black hole with D4 fluxes $p^A$ and D0
potential $\phi^0$.\foot{D2 charges are suppressed here for
brevity and will be restored at the end of this section.}  The $10
d$ string frame metric is \eqn\sedf{ds_{10}^2=ds_{CY}^2+ds_4^2,}
\eqn\tok{ds_4^2=4 \ell^2
\bigl({dr^2+r^2d\theta^2\over(1-r^2)^2}+{1 \over
4}d\Omega_2^2\bigr),} where the $AdS_2$ radius  is\foot{At leading
order $\phi^0 = \pi \sqrt{D_{ABC}p^A p^B p^C \over q_0} $, where
$6 D_{ABC}$ are the Calabi-Yau intersection numbers.}
\eqn\tru{\ell ={g_s }\sqrt{\al}{\phi^0 \over \pi}} The Calabi-Yau
K${\ddot a}$hler class is \eqn\truyp{J={4\pi^3\alpha' \over
\phi^0}p^A } where $\omega_A$ is an integral basis for
$H^2(CY_3)$. The RR field strengths are
\eqn\trip{F^{(4)}=\omega_{S^2}\wedge p^A\omega_A} with the
normalization $\int_{S^2}\omega_{S^2}=1$ and \eqn\gsd{F^{(2)}={2 i
\phi^0 \over \pi}  {2r d\theta \wedge dr \over (1-r^2)^2}.} We
note that $F^{(2)}$ is purely imaginary due to the analytic
continuation to euclidean space, as is the underlying one form
potential \eqn\rsa{A^{(1)}= - {  2 i\phi^0 \over \pi} {d\theta
\over 1-r^2}.}

\subsec{IIA $\to$ M}
In this section we construct the lift of the above euclidean  IIA attractor geometry
to M theory. For constant dilaton the M-theory and IIA metrics are related by
\eqn\uns{ds_{11}^2 =  g_s^2{\al} (d x^{11} +
A^{(1)}_{\mu}dx^{\mu})^2 + d s_{10}^2 } where $x^{11} \sim
x^{11}+2\pi $.  In the case at hand
\eqn\met{ds_{11}^2 =
  ds_{CY}^2 + 4 \ell^2 \left({{dr^2 + r^2 d\theta^2}\over{(1-r^2)^2}} +{1 \over 4} d\Omega^2\right) + g_s^2 \al \left(dx^{11} -{ 2 i \phi^0 \over \pi }  {d\theta\over{1-r^2}}\right)^2 }
The metric is complex because the RR potential $A^{(1)}$ is complex.

Since $F^{(2)}$ is nonzero, the M-theory circle (parameterized by
$x^{11}$) is fibered over $AdS_2$. The metric for this 3D fiber
bundle is \eqn\mads{ds_3^2 = 4 \ell^2 \left[ {{dr^2 + r^2
d\theta^2}\over{(1-r^2)^2}} +({\pi \over \phi^0} dx^{11} -
i{{d\theta}\over{1-r^2}})^2 \right]} This is a complexified
$AdS_3$ quotient. There is also a $G^{(4)}$ flux background coming
from the lift of $F^{(4)}$.

Now let us look at the conformal boundary of this 3D metric, which
is at $r=1$. Conformally rescaling the metric by $(r^2-1)/\ell^2$,
throwing away the normal  $dr^2$ term, writing   $r=1-\epsilon$ and
looking near $\epsilon \rightarrow 0$, we obtain \eqn\bd{ds_{bnd}^2
= d\theta^2 + {2 \pi i \over \phi^0}\, d\theta\, dx^{11} +{\cal
O}(\epsilon)} where $\theta$ and $x^{11}$  are identified modulo
$2\pi$.  The fermion boundary conditions are periodic around the
$x^{11}$ circle but antiperiodic around the $\theta$ circle, as it
is contractible in \mads.

\subsec{Complexified boundary tori}

The metric on the torus with modular paramter $\tau$ is, up to an overall conformal factor

\eqn\stau{ds_{\tau}^2 = |d\theta + \tau d{x^{11}}|^2 = d\theta^2 +
2 Re\tau \,d\theta\, d{x^{11}} + (Re\tau^2 + Im\tau^2)
(d{x^{11}})^2} For periodic (antiperiodic) $x^{11}$ ($\theta$)
boundary conditions, the path integral of a 2D CFT on such a torus
yields the partition function \eqn\ztau{Z(\tau) = Tr_{NS}(-1)^F
q^{L_0} \bar{q}^{\bar{L}_0} = Tr_{NS} (-1)^F e^{2\pi i Re\tau
(L_0-\bar{L}_0)} e^{-2\pi Im\tau (L_0 + \bar{L}_0)}} More
generally, we may wish to compute the above partition function
when $Re \tau$ and $Im \tau$ are not necessarily real, but are
defined by separate analytic continuations to the complex plane.
This can be accomplished by analytic continuation to a complex
metric of the form \stau, but in which the parameters $Re \tau$
and $Im \tau$ are independent complex numbers.\foot{Complex
extrema of the euclidean path integral appear in a  variety of
situations, see \colemanlee\ for a cogent discussion.}

Let us now look at the boundary metric \bd\ in the light of these
comments. Since there is no $(d{x^{11}})^2$ term at $\epsilon=0$,
this corresponds to $(Re \tau)^2 + (Im \tau)^2=0$ or $Im \tau= \pm i
Re \tau$, where the sign is fixed by requiring that the real part of
$Im \tau$ be positive, such that \ztau ~ be nondivergent. Matching
also the coefficient of $d\theta \, d{x^{11}}$ one concludes that
\eqn\imt{ Im \tau = {\pi \over \phi^0} ,~~~ Re \tau = i{\pi \over
\phi^0}} Hence the boundary  CFT partition function on the torus \bd\ simply
computes\foot{In this and subsequent expression we have factored out  the singleton modes which are dual to the
black hole center-of-mass. Hence no $F^2$ insertion is required.}
\eqn\trl{ Z_{IIA}(\phi^0,p^A) = Tr_{NS} (-)^Fe^{-{4\pi^2 \over \phi^0}
 L_0}} Invoking holographic duality of bulk
M-theory to the boundary CFT on the torus, we conclude that the
M-theory partition function on \met\ computes the dual CFT partition
function.  Since there is no $\bar q^{\bar L_0}$ dependence in \trl,
right spectral flow acts trivially and we may finally relate the IIA
partition function to the elliptic genus

\eqn\trl{ Z_{IIA}(\phi^0,\phi^A,p^A) = Tr_{R} (-)^Fe^{-{4\pi^2 \over
\phi^0} L_0 - q_A {\phi^A \over \phi^0}}=Z_{BH}} where we have reinstated the D2 potentials $\phi^A$
conjugate to the D2 charges $q_A$. Indeed, the holographic duality maps Wilson lines for
the Narain lattice of M2-brane charges to the nontrivial $G^{(4)}$
background mentioned in the previous section.

For future reference we record the attractor equations
for the black hole moduli  and the OSV formula for the topological string coupling
$g_{top}$
\eqn\att{ t^A = {\pi p^A + i {\phi^A } \over i {\phi^0
}}~,~~~~~g_{top} = {4 \pi^2 \over \phi^0}}

\subsec{Summary}

  We now put together and summarize the results of this section.
  The string theory partition function on the euclidean D0-D2-D4 attractor geometry lifts to an
M-theory partition on a complexified quotient of $AdS_3\times
S^2\times CY_3$.  Holography then relates this to a path integral of
the dual CFT on the boundary torus, which in turn is identified with
the partition function of the D0-D2-D4 black hole. Putting this all
together and taking into account the fermion boundary conditions, we
conclude that the partition function of IIA string theory on the
euclideanization of the D0-D2-D4 attractor geometry computes the
black hole elliptic genus \eqn\hos{Z_{IIA} (g_{top} = {4 \pi^2 \over
\phi^0} , p^A, \phi^A) = Tr_{R} (-)^Fe^{ -g_{top} L_0 - q_A {\phi^A
\over \phi^0}}} where the trace on the right hand side is over black
hole microstates with center-of-mass factored out.

While we have not done so, we expect it is possible to derive \hos\ without invoking the lift to M-theory.
This might be done along the lines of e.g.  \entropy\  by considering the variation of the
partition function with respect to a deficit angle at the horizon (the center of $AdS_2$)  while being careful about fermion boundary conditions and the complex field strengths. Such a derivation would
have the advantage of extending \hos\ to nonzero D6 charges, for which our  M-theory discussion is
not directly applicable.

\newsec{Worldsheet instantons in attractors}

In this section we derive an expression for the perturbative
string loop expansion of the IIA partition function on an
euclidean attractor, in particular the contribution from
worldsheet instantons.  This calculation could be set up in the
RNS, Green-Schwarz or hybrid formalism.  Each of these formalisms
has a different set of advantages and disadvantages. Here we will
employ the Green-Schwarz formalism as adapted to the study of
worldsheet instantons in \bw\ (which in the end becomes quite
similar to the hybrid formalism \berk).  A great advantage for us
is that the reduction of the Green-Schwarz string to the
topological string was already detailed in \bw, and will not need
to be repeated here.\foot{In \bw certain subtleties concerning
multiple covers in the Green-Schwarz formalism were left
unresolved. While this is an interesting issue in its own right,
it is beyond the scope of the present paper.  Herein we simply
assume an appropriate resolution consistent with the equivalence
of the RNS and Green-Schwarz formalisms.}  At the same time the
AdS spacetime supersymmetries and RR background are easily dealt
with in this formalism.

\subsec{Worldsheet instantons}

Worldsheet instantons wrap a holomorphic cycle $\Sigma$ in the
Calabi-Yau space $CY_3$ and sit at a point in $AdS_2 \times S^2$.
In the Green-Schwarz formalism we employ, the bosonic moduli space
forthe instanton is $AdS_2 \times S^2 \times {\cal M}^Q_g$, where
${\cal M}^Q_g$ is the appropriate moduli space of smoothly
embedded, genus g holomorphic curves in the homology class $Q$ in
the Calabi-Yau threefold.  There are also fermionic zero-modes.
Except for the replacement of $R^4$ by $AdS_2\times S^2$, the
analysis of the Green-Schwarz string in this context is exactly as
in \bw.  We refer to Section 5.2 of \bw\ where it is shown that,
for an instanton that wraps an isolated curve, the internal $CY_3$
degrees of freedom give rise to the correct power of $g_{top}$ and
the correct dependence on the moduli $t^A$ to reproduce the
well-known instanton contribution to the IIA free energy
$F_{top}(t^A)$. Here we take$t^A$ to be at the attractor point. If
the instanton wraps a curve which is neither isolated nor smooth,
the simple analysis of \bw\ must be extended, but we do not
address such technical issues here.  Instead, we concentrate on
the aspects which differ significantly from \bw, namely the zero
modes of the $AdS_2 \times S^2$ bosons and their four goldstino
superpartners, corresponding to the four supersymmetries that are
broken by the worldsheet instanton.  This sector of the theory is
the same at every genus, and gives an overall factor.  In the flat
space computation of \bw\ the corresponding goldstinos have to be
absorbed by the insertion of two graviton vertex operators.  We
will show in the next section that these insertions are
unnecessary in order to obtain a nonzero result in$AdS_2 \times
S^2$, and the spacetime factor multiplies the free energy by a
genus-independent constant.

\subsec{Localization of the $AdS_2\times S^2$ sector}

 In this subsection we compute the integral over the bosonic ($X^\mu$) and fermionic ($\theta^{\a A}$)\foot{ $ \alpha,A
 \in \{1,2\}$, where $\a$ is a negative chirality spinor index and $A$ is an R-symmetry index, see appendix B for details.} zero
 modes which are just constants on $\Sigma$. These are the goldstone bosons and
 goldstinoes associated with the breaking of the $SU(1,1|2)$
 supergroup down to the group generated by the 6 elements conventionally
 denoted $L_0,
 ~~J^3,~~^\pm G^+_{\half},~~^\pm G^-_{-\half}$.\foot{The right
 subscript (superscript) indicates the $J^3$ ($L_0$) eigenvalue,
 while the left superscript denotes the $R$-charge.}.  It can be seen that supersymmetry does not allow quartic fermionic terms for the  zero modes. Hence the action for these zero modes vanishes, and  we get a prefactor
\eqn\rft{Z_4=\int d^4 X d^4\th=0\times \infty.}
 A priori this could be infinity, zero, or a finite number.
 In this section we will use localization to argue that it is
 finite, and then later fix the finite constant by comparison to supergravity.

 The integral \rft\ can be regularized as follows. After gauge-fixing
the local kappa symmetry (see the appendix for details) we are
left with a finite-dimensional group of residual symmetries
associated to the unbroken spacetime supersymmetries.  Of these,
four are linearily realized and four nonlinearly realized on the
worldsheet. The transformation laws are\foot{The $\s^3_A{}^B$
comes from the fact that $\e^1$ is a linear combination of the
$\xi^i$, while $\e^2$ is a linear combination of $\g_5 \xi^i$ (see
appendices A and B2).} \eqn\rgh{\delta_A^i X^{\mu}= 2 \th_A^{\dag}
\g^{\mu} P_+ \xi^i ~, ~~~~~~\delta_A^i \th^B = - \s^3_A{}^B
P_-\xi^i} where $P_{\pm}$ are the positive and respectively
negative chirality projection operators and $\theta^A$ satisfies
$P_+\th^A=0$. Here $\xi^i$ are the four Killing spinors on $AdS_2
\times S^2$, $\xi^i \in \{\xi^{\pm}_{\pm}\} = \xi^{j_3}_{l_0}$.
For an instanton sitting at the north pole of the sphere,
$\d_A{}_+^+$ and $\d_A{}_-^-$ are the linearily realized
supersymmetries, while $\d_A{}_+^-$ and $\d_A{}^+_-$ represent the
broken ones.

Now we replace \rft\ with \eqn\oiy{Z_4=\int d^4 X d^4\theta ~
e^{-tS_{exact}}} where we choose $S_{exact}$ to be the
$\d^-_-$-exact action \eqn\kex{S_{exact} =\delta_{2}{}^-_-
\d_{1}{}^+_+ \left( \e_{\a\b}\, \th_{1}^\a \th_{2}^\b \right)}and
$t$ is a free parameter that we will want to take to infinity. Using
the fact that $ \delta_-^- S_{exact} =0 $ and that the expectation
value of any $\d^-_-$ - exact operator is zero, it is trivial to
show that \oiy\ is independent of the parameter $t$. Therefore \oiy\
is the properly defined version of \rft. The bosonic part of the
action \kex\ is then

\eqn\actn{S_{bos} = (\xi^+_+)^{\dag} P_- \xi^+_+ = {|w|^2 + |z|^2\over(1-|w|^2)
(1+|z|^2)}} So $e^{-t S_{bos}}$ behaves like a gaussian near the north pole of $S^2$ and the origin
of $AdS_2$ and vanishes at the $AdS_2$ boundary $|w|=1$. Therefore

\eqn\intbs{\int d^2 w ~d^2 z \exp\left[-t (|w|^2 + |z^2|)\right]
\sim {{\pi^2}\over{t^2 }} ~~ {\rm as}~~ t\rightarrow \infty} and
localizes the action to $w=z=0$. Thus we have succeeded in
eliminating the infinity from integration over $AdS_2$. Since the
bosonic term has localized the action to $w=z=0$ for $t\to \infty$,
we only need to know \eqn\fmt{S_{f}|_{w=z=0} = 2 \, \sigma^3_{\a\b}
\,\th_1^\a \th_2^\b} It remains to evaluate  the fermionic integral
\eqn\ftint{\int d^4 \th \exp\left[-t S_{ferm}|_{w=z=0} \right] =
\int d^4 \th \,t^2 \, S^2_{ferm}|_{w=z=0}=  8 t^2 \int d^4 \th
\,\th^4 = 8 \, t^2} We see that the leading factors of $t$ in the
fermionic integration cancel the factor of $t$ in the bosonic
integration and there is a finite nonzero $t\to \infty$ limit. Since
the answer is $t$-independent, this limitng value must be exact for
all $t$ and \eqn\rfa{Z_4=finite~~nonzero~~constant.} We have not
kept track of numerical factors of $2, \pi$ etc. in this discussion.

\subsec{Anti-instantons}

As detailed in the appendix, an anti-instanton at the center of
$AdS_2$ and the south pole of the sphere preserves the same
supersymmetry as an instanton at the center of $AdS_2$ and at the
north pole of the sphere. The goldstinoes $\th_A^{\dot \a}$ now
satisfy the opposite chirality condition, but the computation from
the previous section goes through basically unchanged. Now we
obtain \eqn\sbs{S_{bos} = (\xi^+_+)^{\dag} P_+ \xi^+_+ = {1 +
|w^2| |z|^2\over(1-|w|^2) (1+|z|^2)}} which localizes the action
at $w=0, z=\infty$ - that is, at the south pole. Performing the
fermionic intergral then gives a t independent finite result as
for the instanton.

While such anti-instantons preserve the same supersymmetry, they
couple with the opposite sign to the B-field. The action of an
instanton is $t^A q_A$, and that of an anti-instanton ${\bar t}^A
q_A$, where $t^A$ is determined by the attractor equations. Hence
the contribution to the partition function will involve the modulus
$t^A$ replaced by its complex conjugate $\bar t^A$, resulting in a
contribution proportional  $\bar F_{top}$ to the free energy.

\newsec{Conclusion and Summary}

We have seen that the contribution from holomorphic and
antiholomorphic instantons to the IIA free energy in $AdS_2 \times
S^2$ turns out to be\eqn\grw{C\sum d_{g,Q} e^{-t^A q_A}
g_{top}^{2g-2} +c.c=C(F_{top}+\bar F_{top})} where $d_{g,Q}$ are
the Gromov-Witten invariants (assuming a more precise extension of
the results in \bw). We did not compute the overall constant $C$,
which depends on the normalization of the worldsheet functional
measure. This overall non-zero constant can be deduced using the
fact that IIA string theory is described by a supergravity
expansion at low energies.  The easiest term to match is the one
loop $R^2$ term, which contains no powers of $g_s$ and  reduces in
four dimensions to the Euler density. The contribution of this
term to the partition function was evaluated in \msw, using the
fact that the Euler character of $AdS_2\times S^2$ is 2. Agreement
with this result implies $C=1$. Hence, putting together the
results of sections 2 and 3 and exponentiating the single
worldsheet sum to the multi-worldsheet sum, we have finally
that\eqn\osv{|Z_{top}|^2 = Z^{IIA}_{AdS_2 \times
S^2}(p^A,\phi^A,\phi^0)= Z_{BH}(p^A,\phi^A, \phi^0).}

\bigskip

\centerline{\bf Acknowledgements} We are grateful to J. Lapan, W.
Li, J. Marsano, L. Motl, A. Neitzke and J. Seo for helpful
discussions. THis work is supported in part by the DOE grant DE-FG02-91ER40654.

\appendix {A} {Killing spinors in Euclidean $AdS_2 \times S^2$}

We start with the Killing spinor equation in $11 d_L$

\eqn\kill{ \nabla_{\mu} \e +
{{1}\over{288}}(\G_{\mu}{}^{\nu\rho\s\kappa} - 8
\delta_{\mu}{}^{[\nu} \G^{\rho\s\kappa]})F_{\nu\rho\s\kappa}\,\e
=0} where the antisymmetrization procedure satisfies $F_{[p_1 p_2
\ldots p_n]} = F_{p_1 p_2 \ldots p_n}$ for an $n$-form $F$. We
reduce this condition on $S^1\times CY$, where the M-theory $S^1$
is the time circle. The metric and the four-form field strength
are given by \uns and \trip\ respectively, and after decomposing
the Killing spinor to $4 d$ as $\e= \e^1 \otimes \eta_+ + \e^2
\otimes \eta_-$ and a few manipulations, we obtain
\eqn\okil{\nabla_{\mu} \e^1 + {{i}\over{4}} \slf   \g_{\mu} \e^1
=0} where $F = \omega_{S^2}$, and
 $\e^2$ satisfies the same equation, but with a relative minus sign.
 Therefore if $\e^1$ is the solution we get to \okil, the solution
 for $\e^2$ is $\e^2 = \g_{(4)} \e^1$. Let us now work in a convenient coordinate system and find the eight
Killing spinors of $AdS_2 \times S^2$.

We parametrize $AdS_2\times S^2$ by the complex coordinates $z,w$,
with metric

\eqn\zm{ds^2 = {4 dw d{\bar w}\over (1-w{\bar w})^2 } + {4 dz
d\bar{z}\over(1+z\bar{z})^2}} In these coordinates, we have
\eqn\ef{F =  2  {{i dz \wedge d\bar{z}}\over{(1+z\bar{z})^2}} } We
can decompose the $4 d_E$ gamma matrices as \eqn\decm{\g^m=
\sigma^m \otimes \sigma^3_S ~ , ~~~~ \g^i = 1 \otimes \sigma^i ~ ,
~~~~\g_{(4)} = \sigma^3_A \otimes \sigma^3_S} where the $m$
indices stand for $AdS_2$, and $i$ for $S^2$. Then $\slf = 2i
1\otimes \s^3_S$. Writing the Killing spinor equation in these
coordinates we obtain \eqn\kem{(\nabla_m - \half \,\s_m \otimes
I_2 )\e^1 = 0~,~~~~(\nabla_i - \half\, I_2 \otimes \s^3_S \s^i
)\e^1 = 0} from which we see that we can write the solutions as
$\xi^i= \chi_{AdS_2} \otimes \chi_{S^2}$, where $\chi_{AdS_2} $
($\chi_{S^2}$) has to be a Killing spinor of $AdS_2$ ($S^2$).
Since there are two Killing spinors ($\pm$) on $AdS_2$ and two on
the sphere, we get a total of four Killing spinors $\xi^i$, whose
explicit forms are\foot{The superscript denotes the $J_3$
eigenvalue, while the subsript denotes the $L_0$ one.}

\eqn\fdks{ \xi^+_+ = N \, \left(\matrix{ 1 \cr -z \cr w \cr -wz
}\right) ~~~~~ \xi^+_- = N\left(\matrix{ \bar{w} \cr -\bar{w} z
\cr 1 \cr- z }\right) ~~~~~ \xi^-_+ = N\left(\matrix {\bar{z} \cr
1 \cr w \bar{z}\cr w }\right) ~~~~~ \xi^-_- = N\left(\matrix
{\bar{w}\bar{z} \cr \bar{w} \cr \bar{z} \cr 1 }\right) } where $N
= [(1+z\bar{z}) (1-w\bar{w})]^{-\half} $. Then the Killing spinor
$\e^1(w,\bar{w},z,\bar{z})$ can be written as a linear combination
\eqn\cmbs{\e^1(w,\bar{w},z, \bar{z}) = c_i \xi^i = c_1
\cdot\xi^+_+ + c_2 \cdot\xi^+_- + c_3 \cdot\xi^-_+ + c_4
\cdot\xi^-_-} We will see later that the supersymmetries preserved
by an instanton wrapping a holomorphic cycle in the Calabi-Yau
satisfy $ P_- \e^1 = P_- \e^2= 0 $, while those corresponding to
an instanton wrapping an antiholomorphic curve satisfy $P_+
\e^1=P_+ \e^2 = 0$ , where $P_{\pm}$ are the $4 d_E$ chirality
projectors. It is easy to check that these equations are the same
(for any $w$) if we place the instanton at $(z,\bar{z})$, and the
anti-instanton at $(-1/\bar{z}, -1/z)$ (diametrically opposed
points on the sphere). This means that an instanton sitting at a
point $(z,\bar{z})$ and an anti-instanton sitting at the
diametrically opposite point of the sphere preserve the same set
of supersymmetries. Nevertheless, the set of unbroken
supersymmetries changes as we vary the point $z$.

\appendix {B} {Spinors and $\kappa$-symmetry in euclidean space}

\subsec{Majorana conditions in euclidean space}

In usual type IIA theory, the two supersymmetry parameters are
Majorana -Weyl spinors in $10 d_L$, each counting with 16 real
components. Since in $10 d_E$ we cannot impose both these conditions
simultaneously, we choose to impose a Majorana condition on $\e$ -
initially a Dirac spinor with 64 real components - in the case of
euclidean type IIA (since we need two supersymmetries of opposite
chirality) and would choose a Weyl condition for euclidean type IIB.

 Generally speaking, a Majorana condition $\psi^* = B \psi$ can be
 imposed on a spinor $\psi$ if there exists an invertible matrix $B$ such
that $B^{\dag} B =1$, and $B$ is {\it symmetric}\foot{To be very
precise, the action of $B$ on the gamma matrices in $d$ euclidean
dimensions is \spin\ \eqn\defe{\G_m^* = \eta B_{(d)} \G_m
B^{-1}_{(d)}, ~~~~~~\G_{d+1}^* = \G_{(d)}^*= (-1)^{d \over 2}
B_{(d)} \G_{d+1} B^{-1}_{(d)} \;\;\;\;\;\; B_{(d)}^* B_{(d)}= \e ,
\;\;\;\;\;\; \eta^2= \e^2 =1} where the coefficients $\e$ and $\eta$
are to be specified in each dimension. We can also write $B^T = \e B
$. In Polchinski's conventions, $\eta = (-1)^{d \over 2}$ for what
he calls $B_1$, and $\eta= (-1)^{{d \over 2}+1}$ for $B_2$. We can
only impose a Majorana condition on a spinor if $\e=1$ ; the choice
of $\eta$ seems to have more to do with the ability to give a mass
to that spinor. We usually have two choices for $\e,\eta$ in each
dimension: in $10 d_E$, we can choose $\eta = \e = 1 $ or $\eta=\e =
-1$. In $6 d_E$ we have $\e=-\eta = \pm1$, while in $4 d_E$,
$\e=-1$, while $\eta=\pm 1$. We therefore see that it is impossible
to impose a Majorana condition on a single $4 d$ euclidean spinor.
}. We would like to know what are the consequences in $4d_E$ of
imposing a Majorana condition $\Psi^* = B_{(10)} \Psi$ on the $10
d_E$ spinor $\Psi$, which we decompose as \eqn\mgs{\Psi=\psi_1
\otimes \eta_+ + \psi_2 \otimes \eta_-} on ${\cal M}_4 \times CY$,
where $\eta_{\pm}$ are  the covariantly constant spinors (of
definite chirality) on  the Calabi-Yau, satisfying $(\eta_{\pm})^* =
\eta_{\mp}$. Let us decompose the $10 d_E$ gamma matrices into 4 and
6-dimensional pieces as \eqn\tddec{\G^i = \gamma^i \otimes I_6 ~
,~~~~~~ \G^m = \gamma_{(4)} \otimes \gamma^m ~ , ~~~~~~ \G_{(10)} =
\gamma_{(4)} \otimes \gamma_{(6)}} where $i$ denote $4d$ directions
and $m$ the $6d$ ones. If we write $B_{(10)}$ as $B_{(10)} = B_{(4)}
\otimes \gamma_{(6)}$, then we get $\eta =1$ in $4 d_E$, and the
spinors $\psi_1$ and $\psi_2$ get related by a {\it symplectic
Majorana condition}

\eqn\symp{\psi_A^* = - B_{(4)} \e_{AB} \psi^B~, ~~~~~ A=1,2} so we
can always trade $\psi_1^*$ for $\psi_2$ and viceversa, which we
will do in the text. Note also that we have $B_{(4)} \g_{(4)}
B^{-1}_{(4)} = \g^*_{(4)}$, so the subgroups $(2^{k-1})_+$ and
$(2^{k-1})_-$ are left invariant by the action of
$B_{(4)}$.Therefore we can simultaneously impose a Weyl condition on
$\psi_A$, which will be left unchanged by complex
conjugation\foot{That is, complex conjugation does not change the
dotted into the undotted indices that correspond to each of the
$SU(2)$'s in the decomposition $SO(4) = SU(2)\times SU(2)$.}. Our
choice of $B_{(4)}$ will be such that \eqn\bfr{\psi^*_{A\a} =
\e_{\a\b} \e_{AB} \psi^{B\b}~,~~~~~\psi^*_{A{\dot \a}} =  \e_{{\dot
\a}{\dot \b}} \e_{AB} \psi^{B{\dot \b}}}

\subsec{The kappa-symmetry analysis}

The $\kappa$-symmetry variation of the $10 d_E$ (super)coordinates
$X^M$, $\Th$ for the case of a string wrapping a holomorphic cycle
in the Calabi-Yau, can be written as

\eqn\dka{\delta_{\kappa} \Theta  = (1+ \Gamma)\kappa(\sigma)\equiv
2 {\cal P}_+ \kappa(\sigma), ~~~~~~ \delta_{\k} X^M = - \Th^{\dag}
\G^M \d_\k \Th} where $\Gamma$ is given by

\eqn\ksg{\G= {i\over 2}\epsilon^{z{\bar
z}}\partial_{z}X^M\partial_{\bar z}X^N \Gamma_{MN}\Gamma_{11}} On
the other hand, under global supersymmetry these fields transform
as

\eqn\glb{\delta_{\epsilon} \Theta = \epsilon~, ~~~~~~ \d_\e X^M  =
\Th^{\dag} \G^M \e} where $\e$ satisfies the Killing spinor
equation in the background under consideration, in our case $AdS_2
\times S^2 \times CY$, and can be decomposed as in \mgs. Now we
would like to impose the gauge-fixing condition ${\cal P}_+ \Theta
=0$, which makes sense since for any supersymmetry variation
$\delta_{\epsilon} \Theta_+ = \e_+ $, we can always find a
compensating $\kappa$-symmetry transformation $2 \kappa_+ = -
\e_+$ that would keep us in this gauge. (Note that we can always
set $\kappa_- =0$, since such parameters do not contribute to the
change in $\Theta$). This gauge condition is also compatible with
a Majorana condition on $\Theta$, that is \eqn\real{({\cal
P}_+\Theta)^* = B_{(10)} {\cal P}_+\Theta} which is easily checked
using the properties of $B_{(10)}$.

After gauge-fixing, the residual supersymmetries that we are left
with in $10 d_E$ are \eqn\res{\d_\e \Th = {\cal P}_- \e~, ~~~~~~~
\d_\e X^M = 2 \Th^{\dag} \G^M {\cal P}_+ \e}  Now, we decompose
$\Theta$ as \eqn\dcth{\Theta = \th^1 \otimes \eta_+ + \th^2
\otimes \eta_- } where $\th^{1,2}$ are the fermionic superpartners
of the $AdS_2 \times S^2$ coordinates. The $\k$-symmetry condition
on them reduces to $P_+ \th^{1} = P_+ \th^2 =0$, where $P_{\pm}=
\half(1\pm \g_{(4)})$\foot{Generally, $\Th$ decomposes as $\Theta
= \th^1 \otimes \eta_+ + \th^2 \otimes \eta_- + \theta^a \otimes
\g_a \eta_+ + \theta^{\bar{a}} \otimes \g_{\bar{a}} \eta_-$. If we
separate the $a$ directions into tangent and normal to the curve
$\Sigma$ and impose ${\cal P}_+ \Th =0$, then we get the same set
of physical fields as in \bw.}.

Decomposing \res, we get \eqn\fdvar{\d_{\e} \th^A = P_-
\e^A~,~~~~~~ \d_\e X^{\mu} = 2 \th^{\dag}_A \g^{\mu} P_+ \e^A }
where $A$ is summed over $1,2$, and $\e^1 = c_i \xi^i, ~ \e^2 =
d_i \g_{(4)}\xi^i$. Note that $(\xi^{j_3}_{l_0})^*$ can be
rewritten as $B_{(4)} \g_{(4)} \xi^{-j_3}_{-l_0}$, so our Killing
spinors do satisfy the symplectic Majorana condition $\e_1^* =
B_{(4)} \e_2$, where we chose $B_{(4)} = i \s^2 \otimes \s^1$ and
the coefficients $d_i$ are a linear combination of the $c_i^*$. We
can use this Majorana condition for $\th^1$ and $\th^2$ to rewrite
the $X^{\mu}$ variation under each supersymmetry as

\eqn\lisas{ \d_A^i \th^{B{\a}} = - \s^3_A{}^B \xi^{i \a}~,~~~~
\d_A^i X^{\mu} = 2 \th_A^{\dag {\a}} \g^{\mu}_{{\a} \dot\a}
\xi^{i\dot\a} = - 2 \e_{AB} \th^{B{\a}}\g^{\mu}_{{\a} \dot\a}
\xi^{i\dot\a} }

In the case of a string wrapping an antiholomorphic cycle, all
that happens is that $\G$ changes to $-\G$ in \dka. Therefore the
gauge-fixing condition is now ${\cal P}_- \Th =0$ and the analysis
goes basically the same, just changing all $P_+$ into $P_-$ and
viceversa.

\appendix{C}{ The worldsheet fields}

Following \bw\ for the instanton, we need only
describe the physical fields that live on the worldsheet after
fixing the $\kappa$ symmetry.  A key point is that the resulting
fields are automatically twisted if the normal bundle $N$ of
$\Sigma$ in $CY_3$ is nontrivial.

The massless bosons describe fluctuations normal to $\Sigma$ in
$AdS_2 \times S^2 \times CY_3$.  They are valued in the rank four
holomorphic bundle ${\cal O}^2\oplus N$, where ${\cal O}^2$ is the
trivial rank two holomorphic bundle on $\Sigma$ which describes
fluctuations normal to $\Sigma$ inside $AdS_2 \times S^2$ and $N$
is the holomorphic normal bundle to $\Sigma$ inside the Calabi-Yau
threefold.

To describe the physical fermions on $\Sigma$, we must first fix
the kappa symmetry, which we achieve by imposing the gauge
condition \eqn\kap{{\cal P}_+\Theta = 0} Following \bw, we let
$S_+$ be a right-moving spin bundle and $S_-$ be a left-moving
spin bundle associated to the holomorphic tangent bundle of
$\Sigma$.  Before we impose the gauge condition in \kap, the
spinor $\Theta$ transforms as the direct sum of a positive and a
negative chirality spinor in ten dimensions.  In a neighborhood of
the curve $\Sigma$, the ten-dimensional symmetry group $Spin(10)$
reduces to a product $Spin(2) \times Spin(8)$, where $Spin(2)$
acts on the two {\it real} tangent directions along $\Sigma$ and
$Spin(8)$ acts on the eight {\it real} normal directions to
$\Sigma$ in $AdS_2 \times S^2 \times CY_3$.  Under this reduction
from $Spin(10)$ to $Spin(2) \times Spin(8)$, the ten-dimensional
spinor $\Theta$ decomposes into components transforming as

\eqn\compth{\eqalign{&S_+ \otimes S_+({\cal O}^2 \oplus N)\,,\qquad
S_- \otimes S_-({\cal O}^2 \oplus N)\,,\cr &S_- \otimes S_+({\cal
O}^2 \oplus N)\,,\qquad S_+ \otimes S_-({\cal O}^2 \oplus N)\,.}}
Here we naturally use $S_\pm({\cal O}^2 \oplus N)$ to denote the
positive and negative chirality spinors of the normal bundle ${\cal
O}^2 \oplus N$ to $\Sigma$ inside $AdS_2 \times S^2 \times CY_3$.

In terms of the reduction from $Spin(10)$ to $Spin(2) \times
Spin(8)$, we see that the gauge-fixing condition in \kap\ imposes
the constraint that $\Theta$ have definite chirality with respect
to the $Spin(8)$ structure group of the normal bundle ${\cal O}^2
\oplus N$.  In our conventions, the gauge condition \kap\ then
sets the components of $\Theta$ which live in $S_+\otimes
S_-({\cal O}^2 \oplus N)$ and $S_-\otimes S_-({\cal O}^2\oplus N)$
in \compth\ to zero.  So after we fix the kappa-symmetry, the
physical fermions on $\Sigma$ that arise from $\Theta$ transform
as sections of the bundles $S_+ \otimes S_+({\cal O}^2 \oplus N)$
and $S_- \otimes S_+({\cal O}^2 \oplus N)$.  We can further use
the manifest reduction from $Spin(8)$ to $Spin(4) \times Spin(4)$
to write these bundles as

\eqn\sfer{\eqalign{ S_{+}\otimes S_{+}({\cal O}^2 \oplus N) \,&=\,
\Big[S_{+}\otimes S_+({\cal O}^2)\otimes
S_+(N)\Big]\oplus\Big[S_+\otimes S_-({\cal O}^2) \otimes S_-(N)\Big]
\cr S_- \otimes S_+({\cal O}^2 \oplus N) \,&=\, \Big[ S_- \otimes
S_+({\cal O}^2) \otimes S_+(N)\Big] \oplus \Big[ S_- \otimes
S_-({\cal O}^2) \otimes S_-(N)\Big]}}

Thus far, no twisting is evident in \sfer.  As in \bw, the
twisting only becomes apparent when we use the Calabi-Yau
condition to identify the tensor products of the various spin
bundles appearing in \sfer\ as vector bundles on $\Sigma$.  In
particular, we have the isomorphisms \eqn\twst{\matrix{
\eqalign{S_+ \otimes S_-(N) \,&=\, {\cal O} \oplus \wedge^2
N^*\,,\cr S_+ \otimes S_+(N) \,&=\, N^*\,,} &\eqalign{S_- \otimes
S_-(N) \,&=\, {\cal O} \oplus \wedge^2 N\,,\cr S_- \otimes S_+(N)
\,&=\, N\,.}}} Here $N^*$ denotes the conormal bundle, the dual of
$N$.

By convention, the kinetic operator for a right-moving fermion on
$\Sigma$ is a $\partial$ operator (coupled to the appropriate
bundle), and the kinetic operator for a left-moving fermion on
$\Sigma$ is a $\overline\partial$ operator.  Hence it is very
natural to describe the right-moving fermions as transforming as
sections of {\it anti-holomorphic} bundles and the left-moving
fermions as transforming as sections of holomorphic bundles.
Although all the bundles appearing in \twst\ are holomorphic, we
can easily use the holomorphic three-form and the hermitian metric
on the Calabi-Yau to relate these holomorphic bundles to
anti-holomorphic ones.

Hence we identify the twisted right-moving fermions on $\Sigma$ as
transforming as sections of the bundles \eqn\rsec{S_+({\cal
O}^2)\otimes \overline{N}\,,\qquad S_-({\cal O}^2)\otimes
\overline{\cal O}\,,\qquad S_-({\cal O}^2) \otimes
{\overline\Omega}^1_\Sigma} where $\bar{\cal O}$ is the trivial
anti-holomorphic line bundle on $\Sigma$ and
${\overline\Omega}^1_\Sigma$ is the anti-holomorphic bundle of
forms of type $(0,1)$ on $\Sigma$. We denote these right-moving
fermions by $(\bar{\chi}_{\dot \a}^{\bar m}$,$ ~\theta_2^{\a}$,
$\theta_{\bar z}^{\a})$.

Similarly, we see from \twst\ that the twisted left-moving
fermions on $\Sigma$ transform as sections of the holomorphic
bundles

\eqn\lsec{S_+({\cal O}^2)\otimes {N}\,,\qquad S_-({\cal
O}^2)\otimes {\cal O }\,,\qquad S_-({\cal
O}^2)\otimes\Omega^1_\Sigma\,.} We denote these fermions by
$(\chi^m_{\dot\a}$,~$\theta_1^\a$,~$\theta_z^\a)$. As a small
check, we observe that the worldvolume theory on a Type IIA string
must certainly be a non-chiral theory, and our identifications in
\rsec\ and \lsec\ are consistent with this fact.  In \bw, it was
shown that the self-dual part of the graviphoton field strength
couples to the fields $\th_z^{\alpha}$ and $\th_{\bar
z}^{\alpha}$.

For the anti-instanton in physical gauge, one finds from a similar
analysis that the twisted worldvolume fermions transform in nearly
the same bundles on $\Sigma$ as in \rsec\ and \lsec, but with
$S_+({\cal O}^2)$ exchanged for $S_-({\cal O}^2)$ and vice versa.
That is, on the anti-instanton we now have twisted fermions
$(\bar\chi^{\bar m}_\alpha, \theta^{\dot\alpha}_2,
\theta^{\dot\alpha}_{\bar z})$ and $(\chi^m_\alpha,
\theta^{\dot\alpha}_1, \theta^{\dot\alpha}_z)$.  The
anti-self-dual part of the graviphoton field strength couples to
the fields $\th^{\dot\alpha}_{z}$ and $\th^{\dot\alpha}_{\bar z}$.

\listrefs

\end